# A second order thermodynamic perturbation theory for hydrogen bond cooperativity in water

Bennett D. Marshall

*ExxonMobil Research and Engineering, 22777 Springwoods Village Parkway, Spring TX 77389*

## Abstract

It has been extensively demonstrated through first principles quantum mechanics calculations that water exhibits strong hydrogen bond cooperativity. Classical molecular simulation and statistical mechanics methods typically assume pairwise additivity, meaning they cannot account for these 3-body and higher cooperative effects. In this document, we extend second order thermodynamic perturbation theory to correct for hydrogen bond cooperativity in 4 site water. We show that the association theory gives substantially different predictions than the first order result, which does not include hydrogen bond cooperativity. By comparison to spectroscopy, neutron diffraction and molecular simulation data, we show that the theory accurately predicts the hydrogen bonding structure of water as a function of temperature and density.

Bennettd1980@gmail.com

# I: Introduction

It is widely appreciated that a water molecule in the liquid water phase has a substantially larger dipole moment than a water molecule in the gas phase. What is less well known is that water exhibits strong hydrogen bond cooperativity, meaning the energy of a liquid phase hydrogen bond may be substantially stronger than a hydrogen bond in a gas phase dimer. This fact has been extensively demonstrated using first principles quantum calculations.[1–4] For instance, in a tetramer ring with each water molecule bonded at a donor and acceptor site, it has been shown that 25.6 % of the binding energy is due to three body cooperative effects.[2]

While this is widely understood in the quantum chemistry community, there has been little discussion of water hydrogen bond cooperativity in either the classical molecular simulation or applied thermodynamics communities. The most successful approach to treat hydrogen bonding in equations of state is the multi-density statistical mechanics of Wertheim.[5,6] Each bonding state of a molecule is treated as a distinct species and assigned a unique density. This choice allows for the inclusion of steric effects and the natural enforcement of the limited valence of the hydrogen bond attraction. A general solution has been obtained[7] to this approach for an arbitrary number of association sites in first order perturbation theory TPT1. It is this form which provides the hydrogen bond contribution for all statistical associating fluid theory (SAFT) equations of state[8–11].

A fundamental assumption in the development of the multi-density cluster expansion[5] is pairwise additivity of the potential of interaction. Sear and Jackson (SJ) were the first to extend TPT to include hydrogen bond cooperativity for the case of two site chain forming molecules.[12] SJ took an *adhoc* approach, by first developing the theory in the low density limit and then extending to higher densities by decorating the low density results with pair correlation functions.

More recently, Marshall *et al.*[13,14] took a more formal approach by incorporating hydrogen bond cooperativity in terms of higher order perturbation theory. Both the approach of Sear and Jackson as well as Marshall *et al.* were shown to be accurate when compared to molecular simulations.[13]

In this paper, we take the more formal approach to bond cooperativity of Marshall *et al.* and extend it to the case of water which has 4 association sites (2 donor and 2 acceptor). We demonstrate that incorporating bond cooperativity at the TPT2 level allows for the qualitatively correct description of the degree of hydrogen bonding. We then estimate all required parameters to the theory from literature experimental and quantum chemistry results. It is then demonstrated by comparison to spectroscopy, neutron diffraction and molecular simulation data that the theory accurately predicts the hydrogen bonded structure of water as a function of density and temperature.

## II: Theory

In this section, we develop a second order thermodynamic perturbation theory for water which includes the effect of hydrogen bond cooperativity. We consider a 4 site water model which consist of two hydrogen bond acceptor sites ($O_1$, $O_2$) and two hydrogen bond donor sites ($H_1$, $H_2$) in the overall set $\Gamma = \{O_1, O_2, H_1, H_2\}$. We consider water to be a single sphere of diameter $d$, with a hydrogen bond volume between a donor and acceptor site being described by the parameter $\kappa_{OH}$. To include hydrogen bond cooperativity, we consider the following simple model. The first hydrogen bond that a water molecule receives decreases the energy by $\varepsilon_{hb1}$. The energy of the second hydrogen bond will be $\varepsilon_{hb2} > \varepsilon_{hb1}$ if one of the association bonds is incident to a donor site and the other is incident to a receptor site. If both association bonds are incident on sites with the same functionality (two oxygen or two hydrogen sites) the second association bond receives the same energy as the first $\varepsilon_{hb2} = \varepsilon_{hb1}$. This scheme is outlined in Fig. 1, and is consistent with calculations on small water clusters which show that cooperativity between hydrogen bonds is strong when molecules in a trimer are bonded at donor and acceptor sites.[2] We will construct our theory at second order in perturbation which means we can only enforce specific association energies for molecules bonded up to twice. Clusters (all possible trees) with molecules bonded more than twice are then constructed from combinations of the first and second order contributions.

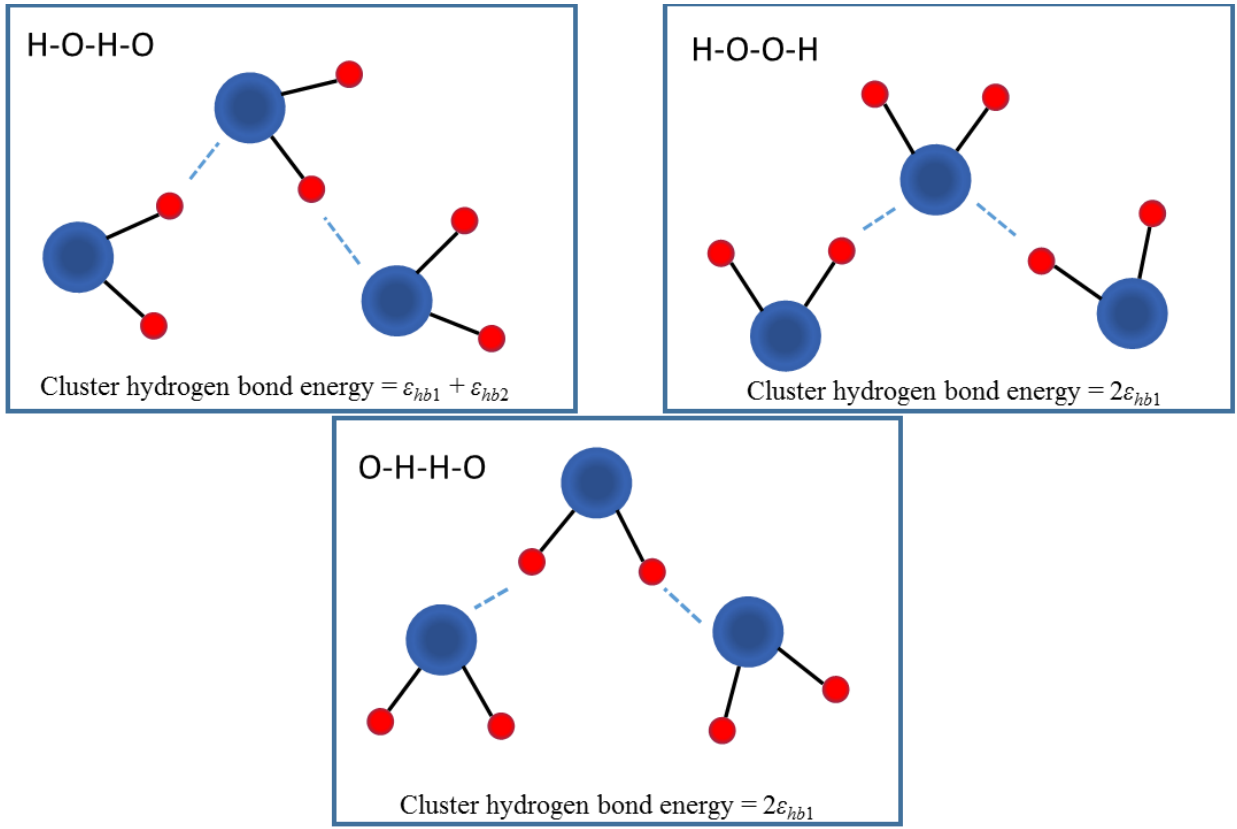

**Figure 1:** Diagrams of associated trimer chains.

## A) Definition of free energy

We develop the theory in the multi-density formalism of Wertheim[5,6,15] where each bonding state of a molecule is assigned a number density. The density of molecules bonded at the set of sites $\alpha$ is given by $\rho_\alpha$. To aid in the reduction to irreducible graphs, Wertheim introduced the density parameters $\sigma_\gamma$

$$\sigma_\gamma = \sum_{\alpha \subset \gamma} \rho_\alpha \qquad (1)$$

where $\gamma$ is a subset of $\Gamma$ and the empty set $\alpha = \varnothing$ is included in the sum. Two notable cases of Eq. (1) are $\sigma_\Gamma = \rho$ and $\sigma_o = \rho_o$; where $\rho$ is the total number density and $\rho_o$ is the density of molecules not bonded at any association site (monomer density). The fraction of molecules NOT bonded at the set of sites $\alpha$ is defined as

$$X_\alpha = \frac{\sigma_{\Gamma-\alpha}}{\rho} \qquad (2)$$

In Wertheim's multi – density formalism the exact change in free energy due to association is given by

$$\frac{A - A_r}{Vk_BT} = \frac{A^{AS}}{Vk_BT} = \rho \ln\left(\frac{\rho_o}{\rho}\right) + Q + \rho - \Delta c^{(o)}/V \qquad (3)$$

where $V$ is the system volume, $T$ is temperature, $A_r$ is the non-associating reference free energy and $Q$ is given by

$$Q = -\rho + \sum_{\substack{\gamma \subset \Gamma \\ \gamma \neq \varnothing}} c_\gamma \sigma_{\Gamma-\gamma} \qquad (4)$$

The term $\Delta c^{(o)}$ is the associative contribution to the fundamental graph sum which encodes all association attractions between the molecules, and $c_\gamma$ is obtained from the relation

$$c_\gamma = \frac{\partial \Delta c^{(o)}/V}{\partial \sigma_{\Gamma-\gamma}} \qquad (5)$$

It was demonstrated by Marshall et al.[13,14] that TPT could be used to enforce cooperative effects in 2 site associating fluids. For the two site case, $\Delta c^{(o)}$ was evaluated in a resummed perturbation expansion over all possible chain lengths. Resummation in the two site case was desired due to the fact each time the linear cluster grew in length, the additional association bond was guaranteed to be cooperative. In the current case for water with 4 sites, as molecules associate into larger clusters, all hydrogen bonds will not exhibit cooperativity. On these grounds, for the current case, we forgo the resummation and opt for a slightly simpler second order approach. Splitting the graph sum into first and second order contributions

$$\Delta c^{(o)} = \Delta c_I + \Delta c_{II} \tag{6}$$

The first order contribution is given by

$$\frac{\Delta c_I}{V} = \frac{1}{2}\sum_{A\in\Gamma}\sum_{B\in\Gamma}\sigma_{\Gamma-A}\sigma_{\Gamma-B}\Delta_{AB} \tag{7}$$

where[16]

$$\Delta_{AB} = \kappa_{AB} g_r d^3 f_{hb1} \tag{8}$$

and $g_r$ is the contact value of the reference system pair correlation function. We define the association Mayer functions to be

$$f_{hbi} = \exp\left(\frac{\varepsilon_{hbi}}{k_b T}\right) - 1 \quad ; \quad i = 1,2 \tag{9}$$

For the second order contribution we use the approach of Marshall and Chapman[13] who showed that higher order perturbation theory could be used to incorporate bond cooperativity in TPT. At second order for the 4 site model this gives

$$\Delta c_{II} = \frac{1}{2}\sum_{A\in\Gamma}\sum_{B\in\Gamma}\sum_{C\in\Gamma}\sum_{D\in\Gamma}\sigma_{\Gamma-A}\sigma_{\Gamma-B}\sigma_{\Gamma-CD}\Delta_{AC}\Delta_{BD}(\delta_{CD}-1) \tag{10}$$

Where

$$\delta_{CD} = \begin{cases} \dfrac{f_{hb2}}{f_{hb1}} & \text{if } CD \in OH \\ 1 & Otherwise \end{cases} \quad (11)$$

Equations (10) and (11) state that a molecule which is bonded at both sites $C$ and $D$ has a non-zero second order contribution if $CD$ are groups of opposite functionality (one $O$ acceptor and one $H$ donor), while if $CD$ are of the same functionality (both $O$ acceptors or both $H$ donors) the second order contribution vanishes because there is no hydrogen bond cooperativity.

**B) Evaluation of bonding fractions**

Now we focus on the self-consistent solution of the relevant densities. Evaluating (5) subject to Eqns. (6), (7) and (10)

$$c_A = \sum_{B \in \Gamma} \sigma_{\Gamma-B} \Delta_{AB} + \sum_{B \in \Gamma} \sum_{C \in \Gamma} \sum_{D \in \Gamma} \sigma_{\Gamma-CD} \sigma_{\Gamma-B} \Delta_{AC} \Delta_{BD} (\delta_{CD} - 1)$$

$$c_{CD} = \sum_{A \in \Gamma} \sum_{B \in \Gamma} \sigma_{\Gamma-A} \sigma_{\Gamma-B} \Delta_{AC} \Delta_{BD} (\delta_{CD} - 1) \quad (12)$$

$$c_\alpha = 0 \quad for \quad n(\alpha) > 2$$

From (6) and (12) we obtain

$$\Delta c^{(o)} = \frac{1}{2} \sum_{A \in \Gamma} \sigma_{\Gamma-A} c_A \quad (13)$$

The densities of the various bonding states can be calculated through the relation[5]

$$\frac{\rho_\gamma}{\rho_o} = \sum_{P(\gamma)=\{\tau\}} \prod_\tau c_\tau \quad (14)$$

From (12) and (14) the density of molecules bonded at site $A$ is given by

$$\rho_{H_1} = \rho_{H_2} = \rho_{O_1} = \rho_{O_2} = \rho_o c_H \quad (15)$$

In Eq. (15) we have enforced that all one site densities are equal. In Eq. (15) for $c_H$ and in the following analysis we omit the subscripts $i$ in the sites ($H_i$) for notational simplicity. For molecules bonded at two association sites, the densities depend on whether both association bonds are of the same functionality as dictated through Eq. (11)

$$\rho_{OH} = \rho_o(c_{OH} + c_O c_H) = \rho_o(c_{OH} + c_H^2)$$

$$\rho_{OO} = \rho_{HH} = \rho_o c_H^2$$
(16)

For molecules bonded three or four times we obtain

$$\rho_{OHO} = \rho_{HOH} = \rho_o(2c_{OH}c_H + c_O c_H^2) = \rho_o(2c_{OH}c_H + c_H^3)$$
(17)

$$\rho_{HOHO} = \rho_o(c_H^2 c_O^2 + 2c_{OH}^2 + 4c_{OH}c_O c_H) = \rho_o(c_H^4 + 2c_{OH}^2 + 4c_{OH}c_H^2)$$

From Eqns. (15) – (17) we construct the densities of molecules bonded $k$ times

$$\rho_1 = 4\rho_H = 4\rho_o c_H \qquad \rho_2 = 2\rho_{HH} + 4\rho_{OH} = \rho_o(6c_H^2 + 4c_{OH})$$

$$\rho_3 = 4\rho_{HOH} = \rho_o(8c_{OH}c_H + 4c_H^3) \qquad \rho_4 = \rho_o(c_H^4 + 2c_{OH}^2 + 4c_{OH}c_H^2)$$
(18)

From Eqns. (18) and the constraint that the sum over all bonding states of a molecule must yield the total density, we solve for the monomer fraction $X_o = \rho_o / \rho$ as

(19)

$$X_o = \frac{1}{c_H^4 + 2c_{OH}^2 + 4c_{OH}c_H^2 + 8c_{OH}c_H + 4c_H^3 + 6c_H^2 + 4c_{OH} + 4c_H + 1} = \frac{1}{(1+c_H)^4 + 4c_{OH}(1+c_H)^2 + 2c_{OH}^2}$$

Now the fraction of molecules NOT bonded at site $H$ is obtained from

$$\rho X_H = \sigma_{\Gamma-H} = \rho_{HOH} + 2\rho_{OH} + \rho_{HH} + 3\rho_H + \rho_o$$
(20)

Using Eqns. (15) – (17) to simplify (20)

(21)

$$\frac{X_H}{X_o} = \frac{\rho_{HOH}}{\rho_o} + 2\frac{\rho_{OH}}{\rho_o} + \frac{\rho_{HH}}{\rho_o} + 3\frac{\rho_H}{\rho_o} + 1 = 2c_{OH}(1+c_H) + (1+c_H)^3$$

The last fraction to consider is the fraction of molecules NOT bonded at both sites $OH$

$$\frac{X_{OH}}{X_o} = \frac{\rho_{OH}}{\rho_o} + 2\frac{\rho_H}{\rho_o} + 1 = c_{OH} + c_H^2 + 2c_H + 1 = c_{OH} + (1+c_H)^2 \qquad (22)$$

Combining Eqns. (19), (21) – (22) we obtain the pair of equations which must be solved to calculate the fractions $X_H$ and $X_{OH}$

$$X_H = \frac{2c_{OH}(1+c_H) + (1+c_H)^3}{(1+c_H)^4 + 4c_{OH}(1+c_H)^2 + 2c_{OH}^2} \qquad (23)$$

and

$$X_{OH} = \frac{c_{OH} + (1+c_H)^2}{(1+c_H)^4 + 4c_{OH}(1+c_H)^2 + 2c_{OH}^2} \qquad (24)$$

with $c_H$ and $c_{OH}$ obtained by simplifying Eqns. (12)

$$c_H = 2\rho X_H \Delta_{OH} + 8\rho^2 X_H X_{OH} \Delta_{OH}^2 (\delta_{OH} - 1)$$

$$c_{OH} = 4\rho^2 X_H^2 \Delta_{OH}^2 (\delta_{OH} - 1) \qquad (25)$$

As can be easily verified, Eqns. (21) – (23) yield the correct first order limit when there is no hydrogen bond cooperativity ($\delta_{OH} = 1$).

## C) Thermodynamic functions

In TPT1 the free energy can be simplified to a form containing only the fraction of unbonded sites $X_A$. This second order theory does not allow for this type of simple representation. Equations (3) – (4), (13) are combined to yield

$$\frac{A^{AS}}{Nk_B T} = \ln X_o + 2c_H X_H + 4c_{OH} X_{OH} \qquad (26)$$

To calculate the association contribution to the chemical potential we use the general relation[5]

$$\frac{\mu^{AS}}{k_b T} = \ln X_o - \frac{\partial \Delta c^{(o)}/V}{\partial \rho} \qquad (27)$$

Where we evaluate the density derivative as

$$\frac{\partial \Delta c^{(o)}/V}{\partial \rho} = \left(2 X_H c_H + 4 X_{OH} c_{OH}\right) \rho \frac{\partial \ln \Delta_{OH}}{\partial \rho} \qquad (28)$$

Now the association contribution to the pressure is evaluated as

$$\frac{P^{AS}}{\rho k_b T} = \frac{\mu^{AS}}{k_b T} - \frac{A^{AS}}{N k_b T} = -\left(2 X_H c_H + 4 X_{OH} c_{OH}\right)\left(1 + \rho \frac{\partial \ln \Delta_{OH}}{\partial \rho}\right) \qquad (29)$$

Equation (29) completes the second order thermodynamic perturbation theory for bond cooperativity in an equation of state for water. Extension to mixtures of water with non-hydrogen bonding molecules is straight forward, however including additional associating species will require additional development.

## III: Qualitative behavior

In this section, we study the general features of the bonding fraction solutions obtained in II. For discussion, we consider the arbitrary water parameter set listed in table 1.

| $d$ (Å) | $\varepsilon_{hb1}/k_b$ (K) | $\kappa_{AB}$ | $R$ |
|---|---|---|---|
| 3 | 1400 | 0.05 | 1.2 |

**Table 1:** Arbitrary parameters for water. Here $R$ is the ratio of second and first order hydrogen bond energies.

Here we define the ratio $R$ to be the ratio of hydrogen bonding energies as

$$R = \varepsilon_{hb2}/\varepsilon_{hb1} \qquad (30)$$

$R = 1$ corresponds to the standard TPT1 approach and $R = 1.2$ means the if a water molecule is bonded at an $O$ group and an $H$ group, the second hydrogen bond is 20% stronger than the first.

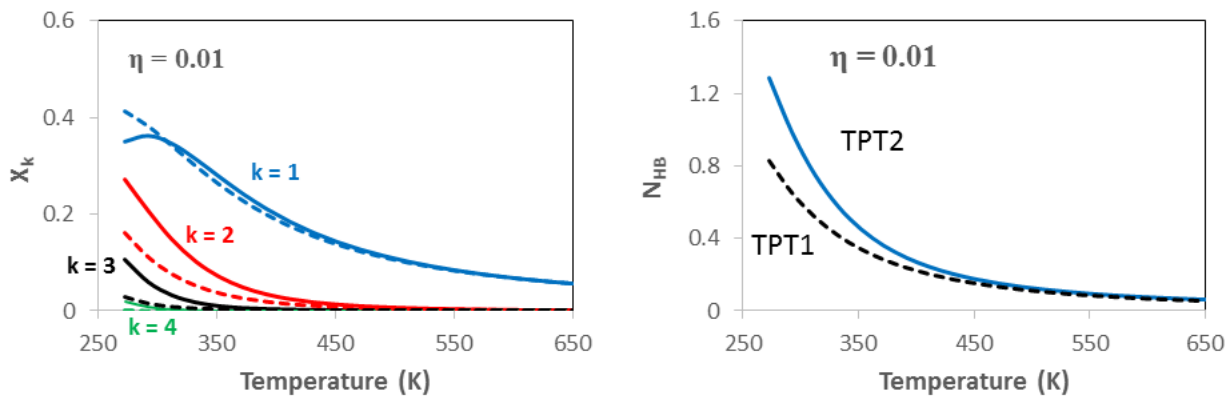

**Figure 2:** Left: Fraction of molecules bonded $k$ times (blue $k = 1$, red $k = 2$, black $k = 3$, green $k = 4$). Solid curves give TPT2 and dashed TPT1. Right: Average number of hydrogen bonds. Both are at $\eta = 0.01$.

This cannot be enforced in TPT1, so we must use the TPT2 approach developed here. Figure 2 plots the fraction of molecules bonded $k$ times $X_k$ versus temperature for a low packing fraction of $\eta = \pi\rho d^3/6 = 0.01$, as well as the average number of hydrogen bonds per water molecule. TPT1 and TPT2 are in good agreement at high temperature when hydrogen bonded molecules typically have a single incident hydrogen bond, however as temperature is decreased and molecules begin to form multiple hydrogen bonds, TPT1 under predicts the degree of hydrogen bonding due to a lack of cooperativity. As can be seen, TPT1 and TPT2 are in good agreement for $X_1$ for temperatures as low as 290 K. It is for the fractions $k > 1$ that TPT1 is in substantial disagreement with TPT2 predictions. Differences between TPT2 and TPT1 are larger at liquid like densities, as can be seen in Fig. 3, which shows the average number of hydrogen bonds at a liquid like packing fraction of $\eta = 0.3$. TPT1 predicts substantially less hydrogen bonding than TPT2 over the entire temperature range.

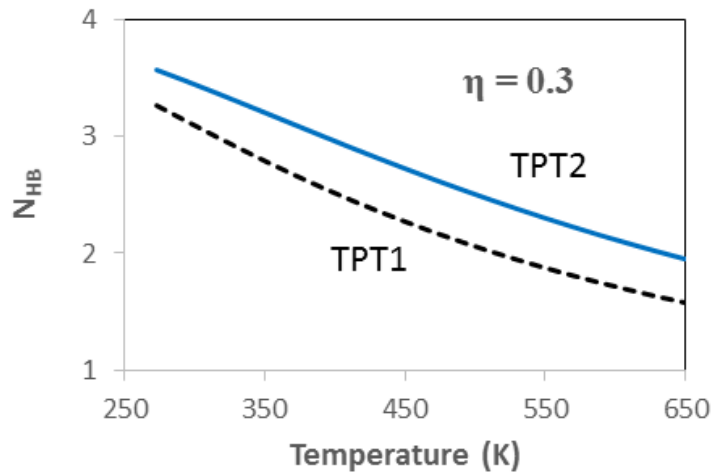

**Figure 3:** Same as right panel Fig. 3 except for $\eta = 0.3$

## IV: Application to water

In this section, we apply the new association theory to predict the hydrogen bonding structure of water. Solution of equations (23) – (24) gives all hydrogen bonding information. These equations explicitly depend on density and temperature. We assume a hard sphere reference fluid with $g_r$ given by the Carnahan and Starling result. Our goal in this section is to estimate the parameters of the model from experimental measurements and quantum mechanical calculations; and then to compare the predictions of the theory to experimental data for the hydrogen bonding structure of water.

Previous studies[9,17,18] using the PC-SAFT equation of state of found that a diameter of $d = 3$ Å works well for water. Here, we keep with this choice and fix the diameter of water to this value. For the association energy $\varepsilon_{hb1}$ we wish to employ known results for the hydrogen bonding strength in the water dimer. The dissociation energy $D_o$ of the water dimer at low $T$ was measured experimentally by Rocher-Casterline et al.[19] to be 13.2 kJ / mol. This value is also in agreement with quantum mechanics calculations for the binding energy of the water dimer above the lowest vibrational state.[20] We take this to be our association energy $\varepsilon_{hb1} / k_b = D_o / R_{ig} = 1587.7\ K$.

To determine the hydrogen bonding energy $\varepsilon_{hb2}$ we use the ratio $R$ defined in Eq. (28) for a hydrogen bonded trimer combined with the hydrogen bond energy $\varepsilon_{hb1}$. Quantum mechanics papers typically report quantities for the lowest energy structure obtained for a cluster size; for a water trimer this is the cyclic configuration.[21] Here we are concerned with trees of hydrogen bonds, as the theory does not account for cyclic structures. To obtain $R$ for a linear trimer we go back to the work of Ojamae and Hermansson[1] who studied the effect of cluster size on cooperativity of water chains and rings. For a trimer chain, they found that the cooperative contribution to the total energy of the cluster was 9%. We can then obtain $\varepsilon_{hb2}$ by the following relation $(\varepsilon_{hb2} + \varepsilon_{hb1})/2 = 1.09\varepsilon_{hb1}$ which yields

$$R = \frac{\varepsilon_{hb2}}{\varepsilon_{hb1}} = 1.18 \tag{31}$$

The final parameter to estimate is the bond volume $\kappa_{OH}$. For this we return to the geometric definition of conical square well association sites,[7] which assumes that if the centers of two molecules (see Fig. 4) are separated by a distance $r < r_c$ and the angles $\theta_A < \theta_{c,A}$ and $\theta_B < \theta_{c,B}$ then the two association sites are considered bonded. For conical square well association sites the bond volume will be given by

$$\kappa_{AB} = \pi(1 - \cos(\theta_{c,A}))(1 - \cos(\theta_{c,B}))\left(\frac{r_c}{d} - 1\right) \tag{32}$$

This specific form is the result of an approximation of the integral of the reference correlation function over the bond volume.[7] Previous experimental studies[22] which measured the average number of hydrogen bonds per water molecule from diffraction data used a similar definition of hydrogen bonding. Soper *et al.*[22] define two water molecules to be hydrogen bonded if the oxygen-oxygen separation in less than 3.5 Å, and the O ··· H − O angle $\upsilon$ is greater than 150°. This corresponds to a critical angle for the hydrogen sites (Fig. 4) of $\theta_{c,H} = 30°$. In the 4 site model, we have split the oxygen site into two distinct association sites. For each of these sites we assume the critical angle $\theta_{c,O}$ to be equal to that of the hydrogen sites. Finally, to evaluate Eq. (32) we must approximate $r_c / d$. The location of the first maximum for the pair correlation function of water in the diffraction data of Soper *et al.*[22] was located at approximately 2.75 Å. If we take this to be the value of their measured diameter, and divide this number into their defined cutoff range of 3.5 Å we obtain $r_c / d = 1.27$. With this information, we evaluate Eq. (32) to obtain $\kappa_{OH} = 0.015$. Table 2 list all association parameters required to evaluate the hydrogen bonding fractions derived in section II.

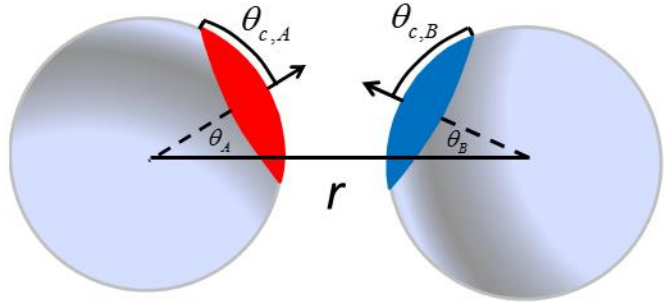

**Figure 4:** Diagram of interacting conical square well association sites

| $d$ (Å) | $\varepsilon_{hb1} / k_b$ (K) | $R$ | $\kappa_{OH}$ |
|---------|-------------------------------|------|---------------|
| 3       | 1587.7                        | 1.18 | 0.015         |

**Table 2:** Association parameters estimated from direct experimental measurement and quantum mechanics calculations

At this point we have not tuned the model to any experimental data. It is desired to compare model predictions to experimental data without tuning. The only way to accomplish this is to compare to hydrogen bonding data as a function of density and temperature. There has been significant discussion in the literature[23,24] on the application of TPT1 to predict Luck's[25] hydrogen bonding data at water saturation. To enforce the "at saturation" condition one must define the reference contribution to the free energy $A_r$ which consist of a hard sphere and spherically symmetric attractive contribution. The introduction of these terms destroys the purity of the comparison of the association contribution of the theory to hydrogen bond structure data. Here, instead of using $T$ and "saturation" as the conditions to evaluate the bonding fractions, we use $T$ and the density which corresponds to saturation at that temperature. The results for the fraction of free OH groups can be found in Fig. 5. The fraction of free OH groups is interpreted as the fractions $X_H = X_O$.[26]

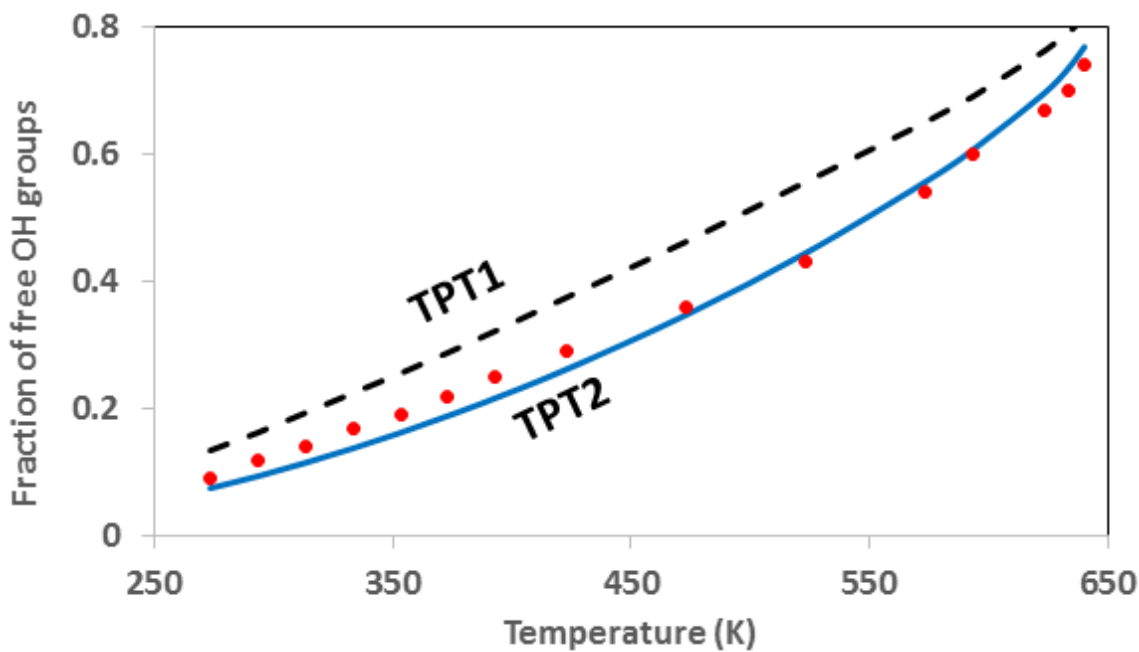

**Figure 5:** Theory predictions (TPT1 – dashed curve, TPT2 – solid curve) and spectroscopy data[25] for the fraction of free OH groups in saturated liquid water. Theory is evaluated as a function of $T$ and saturation density

As can be seen from Fig. 5, TPT2 accurately predicts the hydrogen bonding structure of the fluid as a function of density and temperature. TPT1 under predicts the degree of hydrogen bonding, due to the neglect of hydrogen bond cooperativity. The stunning agreement between theory and experiment gives confidence in the accuracy of the various hydrogen bond fractions developed in section II.

There have been questions as to the accuracy of Luck's data.[23] To sort this out, we compare theoretical predictions for the average number of hydrogen bonds per water $N_{HB}$ to the data of Soper et al.[22], who extracted $N_{HB}$ from pair correlation functions measured from neutron diffraction data. This comparison is made in table 3. Overall model and experimental results are in good agreement. The one point with significant disagreement is for the case $T = 573\ K$ and $\eta = 0.44$. What is particularly encouraging is the good agreement at the ambient temperature of $T = 298\ K$. TPT2 predictions are in good agreement with both Luck's data as well as the $N_{HB}$ data of Soper et al.[22]. This consistency gives a measure of validation to the hydrogen data of Luck.

| $T(K)$ | $\eta$ | $N_{HB}$ (data) | $N_{HB}$ (TPT2) | % error |
|---|---|---|---|---|
| 298 | 0.47 | 3.58 | 3.60 | 0.7 |
| 573 | 0.44 | 3.06 | 2.28 | 25.6 |
| 573 | 0.34 | 1.68 | 1.79 | 6.4 |
| 573 | 0.31 | 1.5 | 1.64 | 9.4 |

**Table 3:** Comparison of TPT2 predictions to experimental data[22] for the average number of hydrogen bonds per water, as a function of temperature and packing fraction $\eta$

We finish this section with a comparison to simulation data for the fraction of molecules bonded $k$ times and $N_{HB}$. Fouad et al.[24] reported simulation results for these quantities at saturation using both the *i*AMOEBA and TIP4P/2005 water models. We compare TPT2 predictions to these simulation results in Fig. 6. As with our previous comparisons, to evaluate the TPT2 hydrogen bonding theory we use the experimental liquid densities at those conditions.

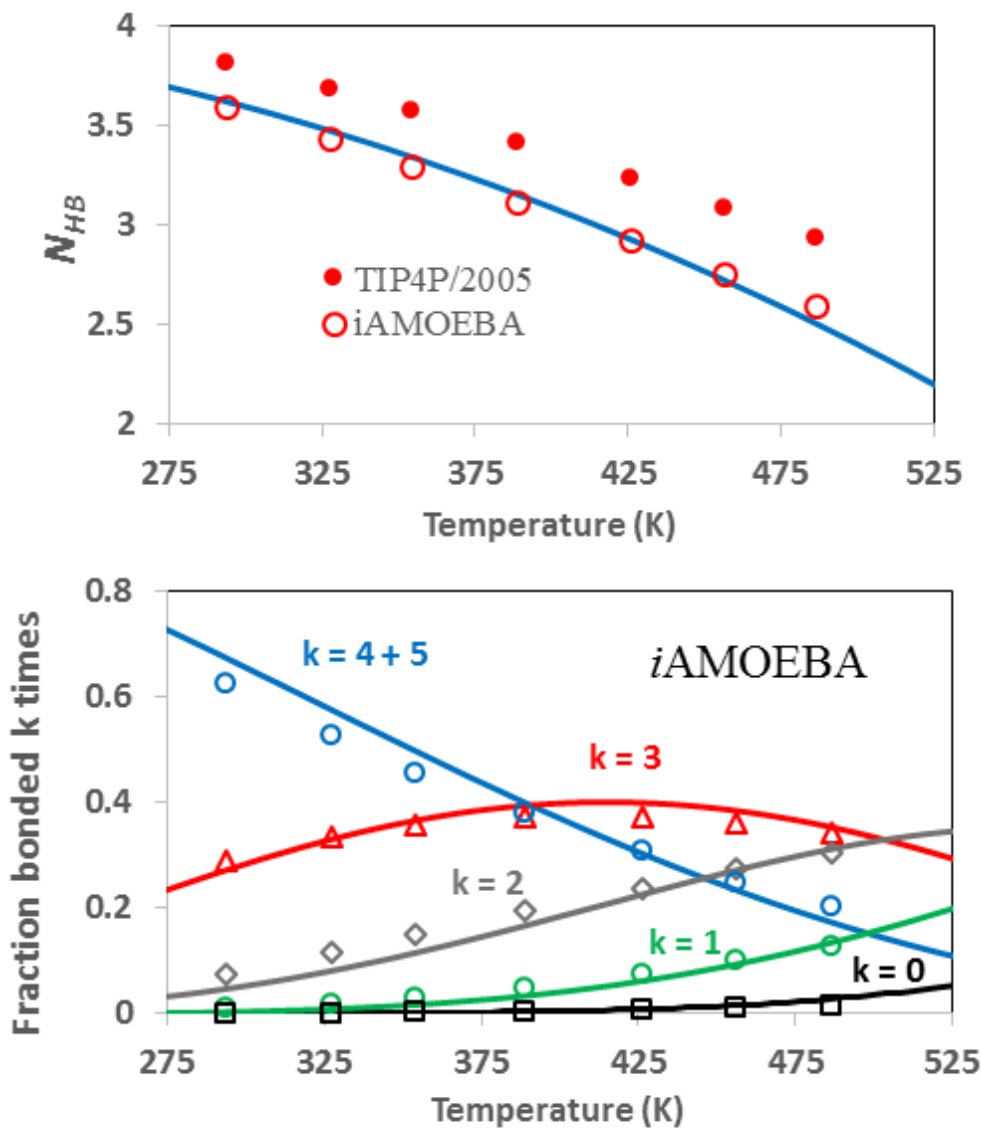

**Figure 6:** TPT2 predictions for hydrogen bond structure compared to the simulation results of Fouad *et al.*[24]. Top panel compares TPT2 (curve) to both *i*AMOEBA and TIP4P/2005 while bottom panel includes comparison between TPT2 (curves) and *i*AMOEBA for the fraction bonded *k* times. TPT2 does not allow for more than *k* = 4 hydrogen bonds per water.

The top panel of Fig. 6 shows results for the average number of hydrogen bonds for TPT2 as well as both simulation models. As can be seen, TPT2 is in very good agreement with the *i*AMOEBA results; the TIP4P results are substantially higher. However, we know from table 3 that TPT2 accurately predicts $N_{HB}$ at ambient conditions. This shows that TIP4P/2005 over predicts the degree of hydrogen bonding, while *i*AMOEBA accurately predicts this quantity. The bottom panel shows the fraction of molecules bonded *k* times as predicted by *i*AMOEBA and TPT2. The agreement between the two methods is remarkable. It should be noted that *i*AMOEBA is a polarizable water model[27], hence both TPT2 and *i*AMOEBA account for 3-body contributions to the internal energy of the system.

## V: Discussion and conclusions

We have extended thermodynamic perturbation theory to account for hydrogen bond cooperativity in a 4 site water model. Using this approach, it was demonstrated that the hydrogen bonding structure of liquid water could be predicted as a function of density and temperature with no tuning to experimental data. TPT2 predictions were in good agreement with Luck's spectroscopy data[25], the neutron diffraction data of Soper *et al.*[22] and classical molecular simulations for water with the *i*AMOEBA force field. TIP4P/2005 was shown to overpredict the degree of hydrogen bonding. This study demonstrates that the hydrogen bonding fractions $X_A$ and $X_{OH}$ calculated through Eqns. (23) – (24) can be used to accurately predict the hydrogen bonding structure of water.

The accuracy of the theory in these predictions gives confidence in the association parameters listed in table 2. Now the question is, how well do these parameters perform in a complete equation of state? To answer this question we employed the PC-SAFT[28] equation of state for the reference free energy in Eq. (3)

$$A_r = A_{HS} + A_{ATT} \tag{33}$$

where $A_{HS}$ is the hard sphere reference and $A_{ATT}$ is the contribution for spherically symmetric attractions. As described previously,[17] we did not include the temperature dependent correction to the hard sphere diameter. $A_{ATT}$ carries a single additional parameter $\varepsilon$ which is the energy scale of attraction. We then attempted to regress a value for $\varepsilon$ by fitting the equation of state to vapor pressure and liquid density data in the temperature range 273.15 K – 583.15 K. We found that an accurate representation of the theory could not be found. Average errors in the vapor pressure were 28% and average errors in liquid density were 7%. To obtain good agreement with experimental data, the bond volume $\kappa_{OH}$ had to be increased from the value of 0.015 listed in table 2. We found that if both $\varepsilon$ and $\kappa_{OH}$ where included in the parameter regression, accurate vapor pressures (3% average error) and liquid densities (2% average error) could be obtained. This resulted in parameter values $\varepsilon = 218.89 k_b$ and $\kappa_{OH} = 0.0564$. As can be seen, the regressed value of $\kappa_{OH}$ is nearly 3.5 times the value of that given in table 2. This value will result in a substantial overprediction in the degree of hydrogen bonding.

On one hand, we have demonstrated that TPT can accurately represent the hydrogen bonding structure as a function of density and temperature. On the other, we have shown that parameters which accurately predict this structure result in an inaccurate equation of state for water. It is the authors belief that the association parameters in table 2 are the correct parameters to use. Optimizations could be performed by adjusting these parameters to the hydrogen bond data; however, they should not change dramatically. The justification for the parameters is their physical basis and the good representation of hydrogen bond structure as a function of density and temperature. The problem then lies in the free energy term $A_r$. Perturbation theories assume that the structure of the fluid (pair correlation function) remains unchanged due to attractive perturbations. For water, this is not the case. At high temperature, water structure will more closely resemble a normal fluid; however, as temperature is decreased and water becomes tetrahedrally coordinated, the density is lower than one would have expected from the high temperature portion of the phase diagram. Hence, if an accurate equation of state for water is to be developed which simultaneously describes the association and phase behavior, this deficiency must be addressed. Note, in $X_A$ and $X_{OH}$ the reference system enters through the reference correlation function $g_r$ in Eq. (8); but $g_r$ is multiplied by $\kappa_{OH}$ which contains a contribution to account for the probability that two hydrogen bonding molecules are correctly oriented for association. This accounting of the orientation dependence, allows for the accurate prediction of bonding fractions even when the assumptions of the perturbation theory have been violated (fluid structure is substantially different than the reference fluid). There is no such orientation dependence in the reference free energy term $A_r$. In a previous publication[17] we demonstrated how the reference energy scale $\varepsilon$ could be coupled to the degree of hydrogen bonding. The last piece of the puzzle is to incorporate communication between the hydrogen bonding and hard sphere reference contributions in both $A_{HS}$ and $A_{ATT}$. This will be the subject of a forthcoming publication.

## References:

[1] L. Ojamaee and K. Hermansson, J. Phys. Chem. **98**, 4271 (1994).

[2] S.S. Xantheas, Chem. Phys. 225 (2000).

[3] E.D. Glendening, J. Phys. Chem. A **109**, 11936 (2005).

[4] U. Góra, W. Cencek, R. Podeszwa, A. van der Avoird, and K. Szalewicz, J. Chem. Phys. **140**, 194101 (2014).

[5] M.S. Wertheim, J. Stat. Phys. **42**, 459 (1986).

[6] M.S. Wertheim, J. Stat. Phys. **42**, 477 (1986).

[7] G. Jackson, W.G. Chapman, and K.E. Gubbins, Mol. Phys. **65**, 1 (1988).

[8] W.G. Chapman, K.E. Gubbins, G. Jackson, and M. Radosz, Ind. Eng. Chem. Res. **29**, 1709 (1990).

[9] J. Gross and G. Sadowski, Ind. Eng. Chem. Res. **41**, 5510 (2002).

[10] F. Llovell and L.F. Vega, J. Phys. Chem. B **110**, 11427 (2006).

[11] A. Gil-Villegas, A. Galindo, P.J. Whitehead, S.J. Mills, G. Jackson, and A.N. Burgess, J. Chem. Phys. **106**, 4168 (1997).

[12] R.P. Sear and G. Jackson, J. Chem. Phys. **105**, 1113 (1996).

[13] B.D. Marshall and W.G. Chapman, J. Chem. Phys. **139**, 214106 (2013).

[14] B.D. Marshall, A. Haghmoradi, and W.G. Chapman, J. Chem. Phys. **140**, 164101 (2014).

[15] M.S. Wertheim, J. Chem. Phys. **87**, 7323 (1987).

[16] W.G. Chapman, G. Jackson, and K.E. Gubbins, Mol. Phys. **65**, 1057 (1988).

[17] B.D. Marshall, J. Chem. Phys. **145**, 204104 (2016).

[18] C.P. Emborsky, K.R. Cox, and W.G. Chapman, Ind. Eng. Chem. Res. **50**, 7791 (2011).

[19] B.E. Rocher-Casterline, L.C. Ch'ng, A.K. Mollner, and H. Reisler, J. Chem. Phys. **134**, 211101 (2011).

[20] X.X. and and I. William A. Goddard, J. Phys. Chem. A **108**, 2305 (2004).

[21] H.M. Lee, S.B. Suh, J.Y. Lee, P. Tarakeshwar, and K.S. Kim, J. Chem. Phys. **112**, 9759 (2000).

[22] A.K. Soper, F. Bruni, and M.A. Ricci, J. Chem. Phys. (1998).

[23] X. Liang, B. Maribo-Mogensen, I. Tsivintzelis, and G.M. Kontogeorgis, Fluid Phase Equilib. **407**, 2 (2016).

[24] W.A. Fouad, L. Wang, A. Haghmoradi, D. Asthagiri, and W.G. Chapman, J. Phys. Chem. B **120**, 3388 (2016).

[25] W.A.P. Luck, Angew. Chemie Int. Ed. English **19**, 28 (1980).

[26] G.M. Kontogeorgis, I. Tsivintzelis, N. von Solms, A. Grenner, D. Bøgh, M. Frost, A. Knage-Rasmussen, and I.G. Economou, Fluid Phase Equilib. **296**, 219 (2010).

[27] L.-P. Wang, T. Head-Gordon, J.W. Ponder, P. Ren, J.D. Chodera, P.K. Eastman, T.J. Martinez, and V.S. Pande, J. Phys. Chem. B **117**, 9956 (2013).

[28] J. Gross and G. Sadowski, Ind. Eng. Chem. Res. **40**, 1244 (2001).